\documentstyle[graphicx,prl,aps]{revtex} 
   \newcommand{\cent}[1] {\begin{center}#1\end{center}}
   \newcommand{\doublint}{\int\rule{-3.5mm}{0mm}\int}
   \newcommand{\mra}  {\to}
   \newcommand{\vecbm}[1]{\mbox{$\boldmath#1$}}

   \newcommand{\lra}  {$\leftrightarrow$}
   \newcommand{\vecb}[1]{\mbox{\bf#1}}
\begin{document}
\paperwidth =15cm
\draft
\title{Micro-canonical Statistical Mechanics of some Non-Extensive
  Systems \footnote{Invited talk at the International Workshop on
    Classical and Quantum Complexity and Non-extensive Thermodynamics,
Denton-Texas,  April 3-6, 2000}}
\author{D.H.E. Gross} \address{ Hahn-Meitner-Institute
  Berlin, Bereich Theoretische Physik,Glienickerstr.100\\ 14109
  Berlin, Germany and Freie Universit{\"a}t Berlin, Fachbereich
  Physik; \today} \maketitle
\begin{abstract} 
  Non-extensive systems do not allow to go to the thermodynamic limit.
  Therefore we have to reformulate statistical mechanics without
  invoking the thermodynamical limit. I.e. we have to go back to
  Pre-Gibbsian times. We show that Boltzmann's mechanical definition
  of entropy $S$ as function of the conserved ``extensive'' variables
  energy $E$, particle number $N$ etc. allows to describe even the
  most sophisticated cases of phase transitions unambiguously for
  ``small'' systems like nuclei, atomic clusters, and selfgravitating
  astrophysical systems: The rich topology of the curvature of
  $S(E,N)$ shows the whole ``Zoo'' of transitions: transitions of
  1.order including the surface tension at phase-separation,
  continuous transitions, critical and multi-critical points. The
  transitions are the ``catastrophes'' of the Laplace transform from
  the ``extensive'' to the ``intensive'' variables. Moreover, this
  classification of phase transitions is much more natural than the
  Yang-Lee criterion.
\end{abstract}

\section{Introduction}
This conference is addressed to the extension of thermo-statistics to
non-extensive systems. This is a new realm of thermo-statistics which
came into focus by the pioneering work of Tsallis \cite{tsallis88}.  
Non-extensive systems are defined by the following property: If they
are divided into pieces, their energy and entropy is not the sum of
the energies and entropies of their parts in contrast to conventional
extensive systems where this is assumed at least if the pieces are
themselves macroscopic. This is the case if the forces in the systems
have a long range comparable with or larger than the linear
dimensions of the system like for nuclei, atomic clusters and
astrophysical objects. However, also inhomogeneous systems, e.g.
systems with separated phases are non-extensive.

Although the largest possible systems like clusters of galaxies belong
to this group I call these systems ``small'' to stress the fact that
the thermodynamic limit either does not exist or makes no sense. 
For systems with short range forces does the entropy of the surfaces
separating the different phases not scale with the volume of the
system. The entropy per particle $s=S(E)/N$ shows a convex intruder
with a depth $\propto N^{-1/3}$.  As long as one cares about this
non-concavity also these systems are to be considered as non-extensive.

Elliot Lieb, Boltzmann laureate from 1998, claims
\cite{lieb97,lieb98a} `` Extensivity is essential for thermodynamics
to work !'' This is certainly true for the original statistical
foundation of thermodynamics by Gibbs \cite{gibbs36}. For the
extension of thermo-statistics to non-extensive, ``small'' systems
one should, however, remember that the original formulation by
Boltzmann, even though he did presumably not think of non-extensive
systems, does not rely on the use of the thermodynamic limit nor any
assumption of extensivity and concavity of the entropy, see below.
Hence, before introducing any major deviation from standard
equilibrium statistics one should explore its original and
fundamental Boltzmann, or micro-canonical, form.

This is what I will do in the following and demonstrate that Lieb's
claim is wrong and contradicts to Boltzmann's view of thermodynamics.
{\em Entropy as defined by Boltzmann does not invoke the thermodynamic
  limit and, consequently, does not demand extensivity.}  Moreover, it
will even turn out that the non-extensivity of inhomogeneous systems
with separated phases gives just a clou to illuminate the physics of
phase transitions explicitly and sharply.  There is a huge world of
non-extensive systems which can only be described by Boltzmann's
micro-canonical ensemble written in its most condensed form on the
epitaph of his gravestone which covers all equilibrium
thermodynamics: \cent{\fbox{\fbox{\boldmath{$S=k*lnW$}}}} where
\begin{eqnarray}
W(E,N,V)&=&\epsilon_0 tr\delta(E-H_N)\label{boltzm}\\
tr\delta(E-H_N)&=&\int{\frac{d^{3N}p\;d^{3N}q}{N!(2\pi\hbar)^{3N}}
\delta(E-H_N)}.\nonumber
\end{eqnarray}
is the volume of the $3N-1$ dimensional manifold at given sharp
energy, the micro-canonical ensemble. ($\epsilon_0$ is a suitable
small energy constant to make the $W$ dimensionless.)

In his famous book ``Elementary Principles in Statistical Physics''
Gibbs deduced the canonical ensemble from the fundamental
micro-canonical. He showed that the canonical one becomes equivalent
to the micro-canonical ensemble {\em in the thermodynamic limit if the
  system is homogeneous}. Otherwise, the canonical and grand-canonical
are not correct. On page 75, chapter VII of \cite{gibbs02} he gives
explicitly the example of the separation of the liquid and gas phase
for which the canonical fluctuations of the energy per particle do not
vanish even in the thermodynamic limit and the canonical ensemble, and
with it the ``Boltzmann-Gibbs'' distribution loose their validity. The
reason for the special fundamental role of the micro ensemble comes of
course from the fact that the internal dynamics of a many-body system
conserves energy and does not mix different energy shells. In an open
system embedded in a heat bath the mechanism of energy violation
operates via the surface between the system and the bath and is of
the same order in the particle number $\propto N^{2/3}$ as any other
internal surface energies, which are also to be ignored in the
canonical treatment.

Boltzmann defines the entropy in eq.(\ref{boltzm}) as a measure of the
mechanical N-body phase space. Thermodynamics has thus a {\em
  geometrical interpretation} and can be read off from the {\em
  topology} of $W(E,N,\cdots)$, the volume of its constant energy
manifold.  No probabilistic interpretation must be invoked like
\begin{equation}
S=-\sum_i{p_i\ln{p_i}},
\end{equation}
where $p_i$ is the probability to find the N-particle configuration
$i$ in the ensemble.

This form of entropy can be investigated for any finite even ``small''
system without any reference to the thermodynamic limit. I will show
here that all thermodynamical features of such a system including the
whole ``zoo'' of phase transitions and critical phenomena can be read
off from this topology. This proves that in contrast to what is
written in most textbooks of statistical mechanics {\em phase
  transitions do exist and can be sharply defined in finite even
  ``small'' and non-extensive systems}.

Conventional thermo-statistics, however, relies heavily on the use of
the thermodynamic limit
($V\!\!\mra\!\!\infty|_{\mbox{\scriptsize{$N/V$, or $\nu$ const.}}}$)
and extensivity, c.f. e.g. the book of Pathria \cite{pathria72}. This
is certainly not allowed for our systems. That the {\em
  micro-canonical} statistics works well also for ``small'' systems
without invoking extensivity will be demonstrated here for finite
normal systems which are also non-extensive at phase transitions of
first order. The use of the thermodynamic limit and of extensivity,
however, is closely intervowen with the development of thermodynamics
and statistical mechanics since its beginning more than hundred years
ago.  When we extend thermodynamics to ``small'' systems we should
establish the formalism of thermodynamics starting from mechanics in
order to remain on a firm basis.  We will see how this idea guides us
to more and deeper insight into the most dramatic phenomena of
thermodynamics, phase transitions.  Moreover, it gives the most
natural extension of thermo-statistics to many non-extensive systems
without invoking any modification of the entropy like that proposed by
Tsallis \cite{tsallis88}. This discussion may further help to
illuminate the domain of physical situations where the Tsallis
formalism is relevant: systems that do not populate the energy
manifold of phase space densely perhaps in a fractal way, perhaps at
the edge of chaos c.f.\cite{latora99}.

I will sketch a deduction of thermo-statistics from the principles of
mechanics alone. Nothing outside of mechanics must be invoked. This
was the starting point of Boltzmann \cite{boltzmann1877} , Gibbs
\cite{gibbs02}, Einstein \cite{einstein03,einstein04} and the
Ehrenfests \cite{ehrenfest12,ehrenfest12a} at the beginning of the
last century.  They all agreed on the logical hierarchy of the
micro-canonical as the most fundamental ensemble from which the
canonical, and grand-canonical ensembles can be deduced under certain
conditions.  According to Gibbs the latter two approximate the micro
ensemble in the thermodynamic limit of infinitely many particles
interacting by short range interactions if the system is homogeneous.
Then surface effects and fluctuations can be ignored relatively to the
bulk mean values. This is the main reason why the thermodynamic limit
became basic in the statistical foundation of macroscopic
thermodynamics.  However, it was Gibbs \cite{gibbs02f} who stressed
that the equivalence of the three ensembles is not true at phase
transitions of first order, even in the thermodynamic limit.

\begin{figure}
\cent{\includegraphics*[bb = 117 13 493 646, angle=-90, width=9cm,  
clip=true]{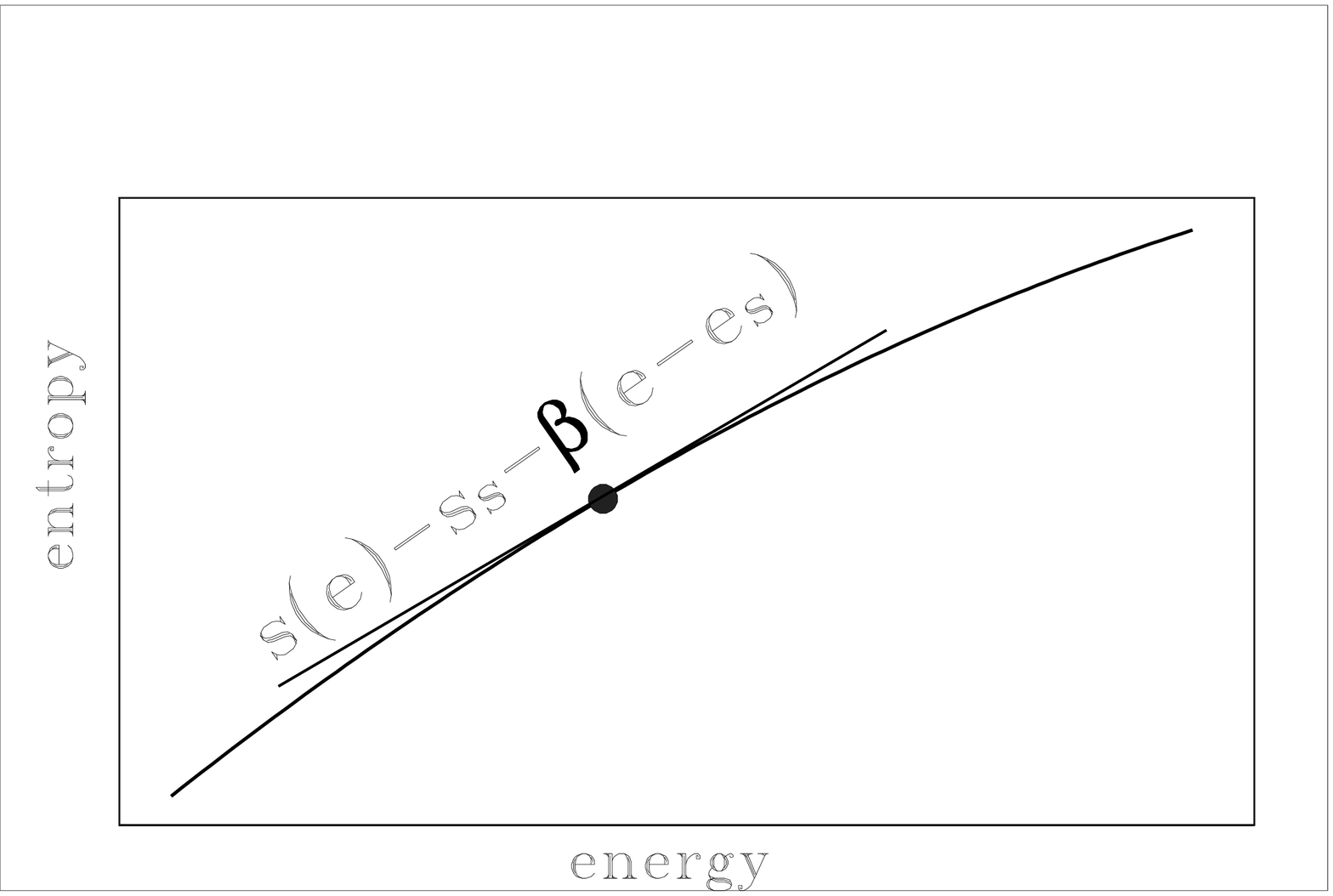}}
\caption{Concave $s(e)$ touching the tangent with slope $\beta$ at a
single stationary point.}\label{concave}
\end{figure}

The link between the micro and the grand ensemble is established by the
double Laplace transform:
\begin{equation} Z(T,\nu,V)=\doublint_0^{\infty}{\frac{dE}{\epsilon_0}\;
dN\;e^{-[E/T+\nu N-S(E)]}}
=\frac{V^2}{\epsilon_0}\doublint_0^{\infty}
{de\;dn\;e^{-V[e/T+\nu n-s(e,n)]}}\label{grandsum}
\end{equation}
Globally $s(e,n)=S(e=E/V,n=N/V)/V$ is concave (downwards bended).  If
$s(e,n)$ is also locally concave then there is a single point
$e_s$,$n_s$ for given $T,\nu$ as shown in figure (\ref{concave}) with
\begin{eqnarray}
\frac{1}{T}=\beta&=&\left.\frac{\partial S}{\partial E}\right|_s\label{beta}\\
\frac{P}{T}&=&\left.\frac{\partial S}{\partial V}\right|_s\label{P}\\
\nu=-\frac{\mu}{T}&=&\left.\frac{\partial S}{\partial N}\right|_s\label{mu},
\end{eqnarray}
and a one to one mapping is generated by the Laplace transform 
eq.(\ref{grandsum}) from the micro variables $E,N$ to the grand
canonical $T,\nu$. This is illustrated by figure (\ref{concave})
for the case of a single conserved variable $E$.
 In Gaussian approximation $s(e)$ remains under the tangent
line and in the thermodynamic limit ($V\to\infty$) only the immediate
neighborhood of the stationary point contributes. The importance of the
curvature of $s(e)$ for the bijective mapping is evident.

An equilibrated many-body system is characterized by few macroscopic
quantities:
\begin{enumerate}
\item Its energy $E$, mass (number of atoms) $N$, volume $V$,
\item its entropy $S$ eq.(\ref{boltzm}),
\item its temperature $T$ eq.(\ref{beta}), pressure $P$ eq.(\ref{P}),
and chemical potential $\nu$ eq.(\ref{mu}).
\end{enumerate}
There are important qualitative differences between these three
groups: All variables of the first group have a clear mechanical
significance. They are conserved and well defined at each point of the
N-body phase space. The internal dynamics of the system cannot leave
the shell in phase space which is defined by these variables.  Also
entropy as the most important quantity within thermodynamics has with
eq.(\ref{boltzm}) a clear mechanical foundation since Boltzmann.  The
set of points on this surface defines the micro-canonical ensemble.
In contrast to the conserved quantities which are defined at each
phase space point, entropy refers to the whole micro-canonical
ensemble.

Remark: For a system with discrete energies $E_i$ with some
degeneration $n(E_i)$ e.g. a lattice or a quantum system one should
define the micro-canonical partition sum by the number $W=n(E_i)$
of states at this energy.  When we discuss derivatives of $W$ we
imagine a suitable smoothing of this.

It is important to notice that Boltzmann's and also Einstein's
formulation allows for {\em defining the entropy entirely within
  mechanics} by $S_{micro}:=ln[W(E,N,V)]$. {\em It is a single valued,
  non-singular, in the classical case multiply differentiable function
  of all ``extensive'', conserved dynamical variables}. No
thermodynamic limit must be invoked and this definition applies to
non-extensive like our ``small'' systems as well.
 
The third group of quantities which characterize the thermodynamical
state of an equilibrated many-body system, temperature $T$, pressure
$P$ and chemical potential $\nu$ have no immediate mechanical
significance. From the mechanical point of view they are secondary,
derived quantities. This difference to the two other groups of
variables will turn out to be significant for ``small'' systems.
Again, like entropy itself, these quantities refer to the whole
micro-canonical ensemble, not to an individual point in the N-body
phase space.

Starting from this point, the conventional thermo-statistics assumes
extensivity and explores the thermodynamic limit
($V\!\!\mra\!\!\infty|_{\mbox{\scriptsize N/V, or $\nu$ const.}}$)
c.f.  \cite{pathria72}. This procedure follows Gibbs \cite{gibbs02}.
He introduced the canonical ensemble, which since then became the basic
of all modern thermo-statistics. 
\section{Phase transitions micro-canonically}
This talk addresses phase transitions in ``small'' systems.
Conventionally phase transitions are thought to exist only in the
thermodynamic limit ($V\!\!\mra\!\!\infty|_{\mbox{\scriptsize N/V, or
    $\nu$ const.}}$). Yang and Lee \cite{lee52} define them by the
singularities of the grand-canonical partition sum as function of the
fugacity $z=e^{\nu}$. As the partition sum $Z(T,\nu)$ is analytical in
$z$ for finite volumes $V$ the singularities can occur in the
thermodynamic limit only ($V\to\infty|_{e,n}$).  I will show how these
singularities arise from points in the parameter space
\{$e,n,\cdots$\} where the micro-canonical (Boltzmann) entropy
$s(e,n)$ has either vanishing or even positive curvature.  These are
the catastrophes of the Laplace transform from the micro-canonical to
the grand-canonical partition sum (\ref{grandsum}).

At phase transitions the inter-phase surface does not scale with the
volume. Systems with phase separation are non-extensive. These
configurations become exponentially suppressed in the canonical
ensemble \cite{gibbs02f}. Moreover, Schr\"odinger thought that
Boltzmann's entropy is not usefull to describe systems other than gases
\cite{schroedinger44}. However, today with the powerfull and cheap
computers we can explore the micro-canonical ensemble in realistic
situation. The clarification the basic role of Boltzmann's statistics
in the most dramatic situations of equilibrium thermodynamics is
demanding.  To realize the powerful application of Boltzmann's [not
Boltzmann-Gibbs (!)] statistics to non-extensive, finite systems
having a non-concave entropy $s(e,n,\cdots)$ may also specify the
cases where one has to go beyond and where a generalization like
Tsallis entropy is needed.

When there are no singularities in the partition sum $Z(T,\nu)$ for
finite systems, are there no phase transitions in finite systems?
There are phenomena observed in finite systems which are typical for
phase transitions. Sometimes this is even so in astonishingly small
systems like nuclei and atomic clusters of $\sim 100$ atoms
\cite{gross95,schmidt97,haberland99}. In chapter (\ref{surftension})
and in reference \cite{gross157} we show that their characteristic
parameters as transition temperature, latent heat, and {\em surface
  tension} are -- in the case of some metals -- already for thousand
atoms close, though of course not equal, to their known bulk values.
Therefore, it seems to be fully justified to speak in these cases of
phase transitions of first order.

We need an extension of thermodynamics to ``small'' systems which
avoids the thermodynamic limit.  However, here is a severe problem:
{\em The non-equivalence of the three popular ensembles, the
  micro-canonical, the canonical, and the grand-canonical ensembles
  for ``small'' systems.} The energy {\em per particle} fluctuates
around its mean value $<\!\!E/N\!\!>$ in the (grand-)canonical
ensemble whereas the energy fluctuations are zero in the
micro-canonical ensemble.  Moreover, the heat capacity is strictly
positive in the canonical ensembles whereas it may become {\em
  negative} in the micro ensemble. To extend themodynamics to
``small'' systems it is certainly advisable to keep close contact with
mechanics.  It is helpful to realize that the fundamental
micro-canonical ensemble as introduced by Boltzmann is the only one
which has a clear mechanical definition
\cite{boltzmann1884,einstein04} for finite systems.

To extend the definition of phase transitions of Yang and Lee to
finite systems we must study which feature of the micro-canonical
partition sum $W(E,N,V)$ leads to singularities of the grand-canonical
potentials $\frac{1}{V}ln[Z]$ as function of $z=e^\nu$ by the Laplace
transform eq.(\ref{grandsum}). In the thermodynamic limit
$V\to\infty|_{\mbox{\scriptsize $\nu$ const.}}$ this integral can be
evaluated by asymptotic methods as discussed above. If there is only a
single stationary point then there is a one to one mapping of the
grand-canonical ensemble to the micro-canonical one and
energy-fluctuations disappear $\propto 1/\sqrt{N}$.

This, however, is not the case at phase transitions of first order.
Here the grand-canonical ensemble contains several Gibbs states
(stationary points, c.f. the figure (\ref{surftens})) at the same
temperature and chemical potential which contribute similarly to the
integral eq.(\ref{grandsum}).  The statistical fluctuations of $e$
and $n$ do not disappear in the grand-canonical ensemble even in the
thermodynamic limit. Between the stationary points {\em $s(e,n)$ has
at least one principal curvature $\geq 0$}.  Here van Hove's
concavity condition \cite{vanhove49} for the entropy $s(e,n)$ is
violated. In the thermodynamic limit these points get jumped over by
the integral (\ref{grandsum}) and $ln[Z]$ becomes non-analytic.
Consequently, we {\em define phase transitions also for finite
systems topologically by the points and regions of non-negative
curvature of the entropy surface $s(e,n)$ as a function of the
mechanical, conserved ``extensive'' quantities like energy, mass,
angular momentum etc.}.

The central quantity of our further discussion, the determinant of the
curvatures of $s(e,n)=S(e=E/V,n=N/V)/V$ is defined as
\begin{equation}
d(e,n)= \left\|\begin{array}{cc}
\frac{\partial^2 s}{\partial e^2}& \frac{\partial^2 s}{\partial n\partial e}\\
\frac{\partial^2 s}{\partial e\partial n}& \frac{\partial^2 s}{\partial n^2}
\end{array}\right\|
= \left\|\begin{array}{cc}
s_{ee}&s_{en}\\
s_{ne}&s_{nn}
\end{array}\right\|=\lambda_1\lambda_2. \end{equation}
The two curvature eigenvalues (main curvatures) are assumed to be
ordered and $\lambda_1>\lambda_2<0$.

Also critical fluctuations, i.e. abnormally large fluctuations of some
extensive variable in the grand-canonical ensemble or the eventual
divergence of some susceptibilities are micro-canonically connected to
the vanishing of the curvature determinant, e.g. in the following
examples of $d(e,n)$ or $d(e,m)$ respectively:
\begin{eqnarray}
&&\mbox{The micro-canonical specific heat is given by :}\nonumber\\
c_{micro}(e,n)&=&\left.\frac{\partial e}{\partial T}\right|_\nu
=-\frac{s_{nn}}{T^2d(e,n)}\;,\\
d&=&\frac{d (\beta\nu)}{d(en)}\\
&&\mbox{or the isothermal magnetic susceptibility by :}\nonumber\\
\chi_{micro,T}(e,m)&=&\left.\frac{\partial m}{\partial B}\right|_T
=\frac{s_{ee}}{d(e,m)}\;,\\
&&\mbox{with }s_{ee}=\frac{\partial^2 s}{\partial e\partial e}\mbox{ etc.}
\end{eqnarray}
In the case of a classical continuous system $s(e,n)$ is everywhere
finite and multiply differentiable. In that case the inverse
susceptibilities $[c_{micro}(e,n,V)]^{-1}$ and $[\chi_{micro,
  T}(e,m,V)]^{-1}$ are well behaved smooth functions of their
arguments even at phase transitions.  Problems arise only if the
susceptibilities are considered as functions of the ``intensive''
variables $T$, and $\nu$ or $B\propto\partial S/\partial m$
\cite{straley73}.  In the case of lattice systems we can only assume
that the inverse susceptibilities are similarly well behaved.

Experimentally one identifies phase transitions of first order of
course not by the non-analyticities of $\frac{1}{V}ln[Z]$ but by the
interfaces separating coexisting phases, e.g. liquid and gas, i.e. by
the {\em inhomogeneities} of the system which become suppressed in the
thermodynamic limit in the grand-canonical ensemble. This fact was
early realized by Gibbs \cite{gibbs06} and he emphasized that using
$S$ vs.  volume, or density, at phase separation ``has a substantial
advantage over any other method (e.g. pressure) because it shows the
region of simultaneous coexistence of the vapor, liquid, and solid
phases of a substance, a region which reduces to a point in the more
usual pressure-temperature plane.''  That is also the reason why for
the grand-canonical ensemble the more mathematical definition of phase
transitions \cite{lee52} is needed.  The main advantage of the
micro-canonical ensemble is that it allows for {\em inhomogeneities}
as well and thus we can keep much closer to the experimental criteria
for finding phase transitions.

Interfaces have three opposing effects on the entropy:
\begin{itemize}
\item An entropic {\em gain} by putting a part ($N_1$) of the system
  from the majority phase (e.g. solid) into the minority phase
  (bubbles, e.g. gas) with a higher entropy per particle. However,
  this has to be paid by additional energy $\Delta E$ to break the
  bonds in the ``gas''-phase. As both effects are proportional to the
  number of particles $N_1$ being converted, this part of the entropy
  rises linearly with the additional energy.
\item With rising size of the bubbles their surfaces grow. This is
  connected to an entropic {\em loss} due to additional correlations
  between the particles at the interface(surface entropy) proportional
  to the interface area. As the number of surface atoms is $\propto
  N_1^{2/3}$ this is not linear in $\Delta E$ and leads to a convex
  intruder in $S(E,N,V)$, the origin of surface tension
  \cite{gross150}. This is also the reason why systems with phase
  separation are non-extensive c.f. chapter (\ref{secondlaw}).\label{surf}
\item An additional mixing entropy for distributing the
  $N_1$-particles in various ways over the bubbles.
\end{itemize}
At a (multi-) critical point two (or more) phases become
indistinguishable because the interface entropy (surface tension)
and with it the inhomogeneity (interface) disappears.

In order to demonstrate this we investigate in the following the
3-states diluted Potts model now on a {\em finite} 2-dim (here
$L^2=50^2$) lattice with periodic boundaries in order to minimize
effects of the external surfaces of the system. The model is defined
by the Hamiltonian:
\begin{eqnarray}
H&=&-\sum_{i,j}^{n.n.pairs}o_i o_j\delta_{\sigma_i,\sigma_j}\\
n&=&L^{-2}N=L^{-2}\sum_io_i .\nonumber
\end{eqnarray}
Each lattice site $i$ is either occupied by a particle with spin
$\sigma_i =1,2,\mbox{ or }3$ or empty (vacancy).  The sum is over
neighboring lattice sites $i,j$, and the occupation numbers are:
\begin{equation}
o_i=\left\{\begin{array}{cl}
1&\mbox{, spin particle in site }i\\
0&\mbox{, vacancy in site }i\\
\end{array}\right. .
\end{equation}

This model is an extension of the ordinary ($q=3$)-Potts model to
allow also for vacancies. At zero concentration of vacancies ($n=1$),
the system has in the limit of infinite volume $V$ a continuous phase
transition at $e_c=1+\frac{1}{\sqrt{q}}\approx
1.58$\cite{baxter73,pathria72}.  With rising number of vacancies the
probability to find a pair of particles at neighboring sites with the
same spin orientation decreases.  The inclusion of vacancies has the
effect of an increasing effective $q_{eff}\ge 3$. This results in an
increase of the critical energy of the continuous phase transition
with decreasing $n$ and provides a line of continuous transition,
which is supposed to terminate when $q_{eff}$ becomes larger than $4$,
where the transition becomes first order.

At smaller energies the system is in one of three ordered phases
(spins predominantly parallel in one of the three possible
directions).  We call this the ``solid'' phase. This scenario gets
full support by our numerical findings.

\begin{figure}
  \includegraphics*[bb =0 0 290 180, angle=-0, width=13cm,
clip=true]{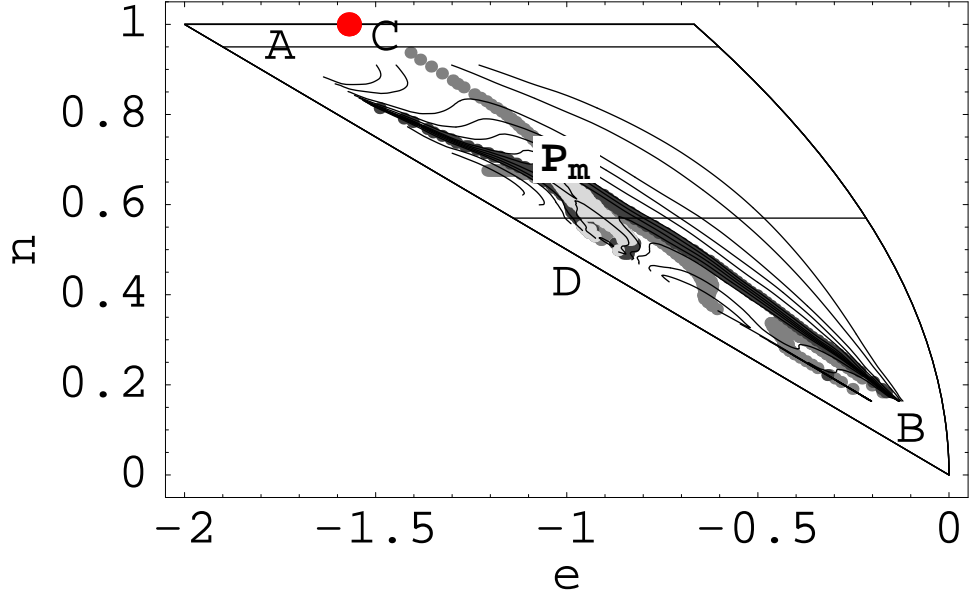}
\\
\caption{Contour plot of the determinant of curvatures
  $d(e,n)$ defined in eq.(\ref{curvdet}).The grey/black strips separate:
  regions above $\widehat{CP_mB}$ : concave, $d>0$, pure phase
  (disordered, gas), in the triangle $A$$P_m$$C$ concave, pure phase
  (ordered, solid); below $\widehat{AP_mB}$: convex, $d<0$,
  phase-separation, first order; at the dark lines $\widehat{AP_mB}$
  we have $d(e,n)=0$: termination or border lines of the first order
  transition; medium dark lines e.g $\widehat{CP_m}$.:
  $\vecb{v}_1\cdot\mbox{\boldmath$\nabla$}d=0$; here the curvature
  determinant has a minimum in the direction of the largest curvature
  eigenvector $\vecb{v}_1$; in the cross-region (light grey) we have:
  $d=0\wedge\mbox{\boldmath$\nabla$}d=\mbox{\boldmath$0$}$ this is the
  locus of the multi-critical point $P_m$ where the larger curvature
  $\lambda_1\equiv 0$ and $s(e,n)$ is cylindrical up to at least third order
  in $\Delta e$ and $\Delta n$. The two horizontal lines give the
  positions of the two cuts shown in figs.\ref{cut1},\ref{cut2}.
\label{det}}
\end{figure}

In figure (\ref{det}) the determinant of curvatures of $s(e,n)$:
\begin{equation}
d(e,n)= \left\|\begin{array}{cc}
\frac{\partial^2 s}{\partial e^2}& \frac{\partial^2 s}{\partial n\partial e}\\
\frac{\partial^2 s}{\partial e\partial n}& \frac{\partial^2 s}{\partial n^2}
\end{array}\right\|
= \left\|\begin{array}{cc}
s_{ee}&s_{en}\\
s_{ne}&s_{nn}
\end{array}\right\|=\lambda_1\lambda_2
 \label{curvdet}
\end{equation}
is shown. On the diagonal we have the ground-state of the $2$-dim
Potts lattice-gas with $e_0=-2n$, the upper-right end is the complete
random configuration (here without contour lines), with the maximum
allowed excitation $e_{rand}=-\frac{2n^2}{q}$ and the maximum possible
entropy. In the region above the line $\widehat{CP_mB}$ we have the
disordered, ``gas''.  Here the entropy $s(e,n)$ is concave ($d>0$),
both curvatures are negative (we have always the smaller one
$\lambda_2<0$).  This is also the case inside the triangle $A$$P_m$$C$
(ordered, ``solid'' phase). In these regions the Laplace integral
eq.(\ref{grandsum}) has a single stationary point. They correspond to
pure phases. 

\begin{minipage}[t]{7cm}
\begin{figure}
  \includegraphics*[bb =0 0 290 180, angle=-0, width=7cm,
  clip=true]{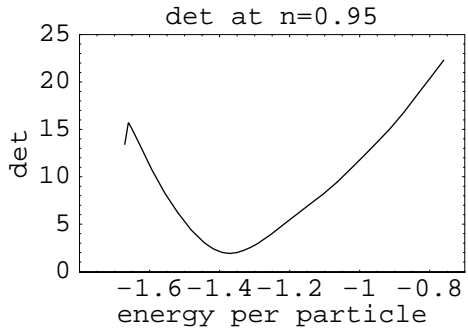}
  \\
\caption{Cut through the determinant $d(e,n)$ along 
  the upper line shown in figure (\ref{det}) at const.  $n=0.95$,
  through the critical line $\widehat{CP_m}$ close to the critical
  point $C$ of the ordinary Potts model ($n\sim 1$)
\label{cut1}}
\end{figure}
\end{minipage}\rule{0.5cm}{0mm}\begin{minipage}[t]{7cm}
\begin{figure}
  \includegraphics*[bb =0 0 290 180, angle=-0, width=7cm,
  clip=true]{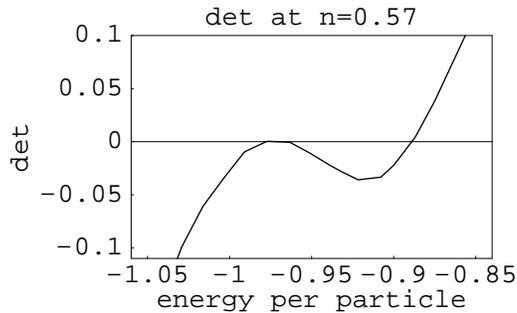}
  \\
\caption{Cut through the determinant $d(e,n)$ along 
  the line shown in figure (\ref{det}) at const.  $n=0.57$, slightly
  below the multi-critical region. There are several zero points of
  the determinant of curvatures: The left one is simultaneously a
  maximum with $\mbox{\boldmath$\nabla d=0$}$ and consequently
  critical as discussed above 
 \label{cut2}}
\end{figure}
\end{minipage}
Below $\widehat{AP_mB}$ $s(e,n)$ is convex ($d<0$) corresponding to
phase-separation, first order.  At these \{$e,n$\} the Laplace
integral (\ref{grandsum}) has no stationary point. Here we have a
separation into coexisting phases, e.g. solid and gas. Due to the
inter-phase surface tension or the negative contribution to the
entropy by the additional correlations at the phase boundaries
(surfaces), $s(e,n)$ has a {\em convex} intruder with positive largest
curvature.  In \cite{binder78,binder82,gross150,gross157} it is shown
that the depth of the convex intruder in $s(e,n)$ gives the surface
tension, c.f. chapter (\ref{surftension}). At the dark lines like
$\widehat{AP_mB}$ we have $d(e,n)=0$.  These are the termination
lines of the first order transition. At these lines one of the two
phases is depleted and beyond all particles are in the other phase
(solid or gas respectively).

Along the medium dark lines like $\widehat{P_mC}$ we have
$\vecb{v}_1\cdot\mbox{\boldmath$\nabla$}d=0$, here the curvature
determinant has a minimum in the direction of the largest curvature
eigenvector $\vecb{v}_1$. The line $\widehat{P_mC}$ towards the
critical point of the ordinary ($q=3$)-Potts model at $e=-1.58$, $n=1$
correponds to a critical line of second order transition which
terminates at the multicritical ``point'' $P_m$.  It is {\em a deep
  valley} in $d(e,n)$ c.f. fig.\ref{cut1} which rises slightly up
towards $C$. On the level of the present simulation we cannot decide
whether this rise is due to our still finite, though otherwise
sufficient, precision or is a general feature of finite size. (The
largest curvature $\lambda_1$ of $s(e,n)$ has a local {\em maximum}
with $\lambda_1\stackrel{<}{\sim}0$, or $d\stackrel{>}{\sim}0$).
Because of our finite interpolation width of $\Delta e\sim \pm 0.04$,
$\Delta n\sim\pm 0.02$ it might be that this valley of $d(e,n)$ gets a
little bit filled up from its sides and the minimum of $d(e,n)$ is
rounded, c.f. fig.\ref{cut1}.  The valley converts below the crossing
point $P_m$ into {\em a flat ridge} inside the convex intruder of the
first order lattice-gas phase-separation region see e.g.
fig.(\ref{cut2}).

In the cross-region (light grey in fig.\ref{det}) we have:
$d=0\wedge\mbox{\boldmath$\nabla$}d=\mbox{\boldmath$0$}$.  This is the
locus of the multi-critical point $P_m$ where the large curvature
$\lambda_1\equiv 0$ in a {\em two-dimensional} neighborhood. Here the
curvature determinant $d(e,n)$ is flat up to at least second order in
both directions $\Delta e$ and $\Delta n$ and $s(e,n)$ is cylindrical.
It is at $e_m\sim -1$, $n_m\sim 0.6$ or $\beta_m= 1.48\pm 0.03$,
$\nu_m= 2.67\pm 0.02$.  Naturally, $P_m$ spans a much broader region
in \{$e,n$\} than in \{$\beta,\nu$\}, remember $d(e,n)$ is {\em flat}
near $P_m$. This situation reminds very much the well known phase
diagram of a $^3$He --$^4$He mixture in temperature vs.  mole fraction
of $^3$He c.f.  fig.3. in ref.\cite{lawrie84}.

\section{On the topology of curvatures}
\begin{eqnarray}
&&\mbox{The two eigenvalues of the curvature matrix (\ref{curvdet}) are:}
\nonumber\\
\lambda_{1,2}&=&\frac{s_{ee}+s_{nn}}{2}\pm\frac{1}{2}\sqrt{(s_{ee}
+s_{nn})^2-4d}\\
&&\mbox{and the corresponding eigenvectors are :}\nonumber\\
\vecb{v}_\lambda&=&\frac{1}{\sqrt{(s_{ee}-\lambda)^2+s_{en}^2}}
\left(\begin{array}{l}
                                        -s_{en}\\
                                         s_{ee}-\lambda
                                    \end{array}\right).\\
\end{eqnarray}
At critical points the following conditions hold:
\begin{eqnarray}
d&=&-\frac{\partial(\beta\nu)}{\partial (en)}=L^2D=0\\
s_{ee}s_{nn}&=&s_{en}^2.\\
&&\mbox{Here the directions $\beta=$const. and $\nu=$ const. are {\em
parallel,}}\\
&&\mbox{the Jacobian vanishes and we have :}\nonumber\\
\left.\frac{\partial \beta}{\partial e}\right|_{\nu}&=&\frac{d}{s_{nn}}=0\\
\left.\frac{\partial \nu}{\partial n}\right|_{\beta}&=&\frac{-d}{s_{ee}}=0.\\
\lambda_1&=&0\\
\lambda_2&=&s_{ee}+s_{nn}\\
\vecb{v}_{\lambda=0}&=&\frac{1}{\sqrt{s_{ee}^2+s_{en}^2}}\left(\begin{array}{l}
                                        -s_{en}\\
                                         s_{ee}
                                    \end{array}\right)\\
\vecb{v}_{\lambda\le0}&=&\frac{1}{\sqrt{s_{nn}^2+s_{en}^2}}\left(
\begin{array}{l}
s_{en}\\ s_{nn} \end{array}\right).\end{eqnarray}

The vanishing of $d$ alone is not sufficient for criticality.
Physically, it means that the surface entropy (tension) and with it
the interface separating coexistent phases disappears. This, however,
can also signalize a depletion of one of the two phases in favor of
the other.  At a critical end-point, however, the interface disappears
at a non vanishing number of atoms in each of the two phases.  I.e.
in an infinitesimal neighborhood of a critical point, $d$ must remain
zero.  In a topologically formulation a critical end-point of first
order transition is at:
\begin{eqnarray}
                                   d&=&0\\
                                   &\mbox{\em and}&\nonumber\\
                                   \vecb{v}_1\cdot\mbox{\boldmath$\nabla$}
                                   d&=&0 ,
\end{eqnarray} 
whereas at a multi-critical point we have {\boldmath$\nabla d=0$}.

This is a generalization of the well known condition for a continuous
transition in one dimension: the simultaneous vanishing of
$\beta^\prime(e)=0$ and of the curvature of $\beta(e)$,
$\beta^{\prime\prime}(e)=0$.

\begin{center}\begin{figure}
    \includegraphics*[bb =0 0 290 180, angle=-0,
    width=11cm,clip=true]{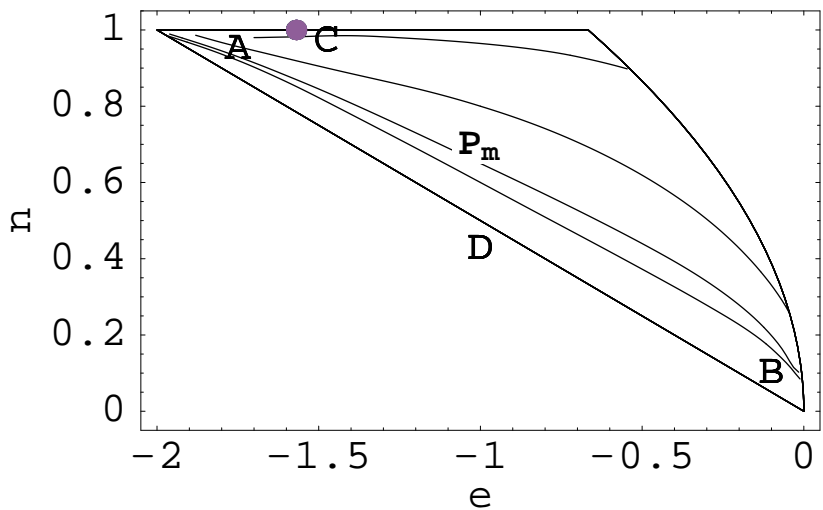}
\caption{Direction of the largest principal curvature $\mbox{\boldmath$v_1$}$ .
\label{maincurvature}}
\end{figure}\end{center}

Figure (\ref{maincurvature}) shows a map of some trajectories which
follow the eigen-vector $\mbox{\boldmath$v_1$}$ with the largest
curvature eigen-value $\lambda_1$. In the region of the convex
intruder ($\lambda_1>0$) i.e. the region of phase-separation
$\mbox{\boldmath$v_1$}$ is $\sim$ parallel to the ground state
$e=-2n$. Also the lines of $\beta=$const. and $\nu=$const. follow
approximately this direction. Their Jacobian
$\partial(\beta\nu)/\partial(en)=d(e,n)$ is negative but small.  This
reminds of the situation in the thermodynamic limit where this region
of phase separation is flat ($d(e,n)$) or $s(e,n)$ cylindrical, both
intensive variables are constant and the Jacobian $d\to-0$.  One can
also see in fig.\ref{maincurvature} how the direction of the largest
curvature $\mbox{\boldmath$v_1$}$ turns into the $e$-direction when
one approaches the critical point $C$ of the ordinary ($q=3$)-Potts
model at $n=1$.

At $n=1$ we know that for an infinite system the ordinary ($n=1$)
three state Potts model has a second order transition at $e=-1.58$
where the curvature of $s(e)$ vanishes, $s_{ee}=0$. I.e. the component
$\vecb{v}_1\cdot\mbox{\boldmath$\nabla$} d$ of
$\mbox{\boldmath$\nabla$} d$ indicates nicely the locus of the second
order ``temperature driven'' transition of the ordinary Potts model.

This chapter was an overview on the power and extreme rich insight
that the topology of curvatures of Boltzmann's micro-canonical entropy
$s(e,n)$ can give in a generic case of a ``small'' and non-extensive
system.
\section{What is the physics behind a positive curvature? }
\label{surftension}
\begin{figure}
\includegraphics*[bb = 84 43 409 303, angle=-0, width=12cm,  
clip=tru]{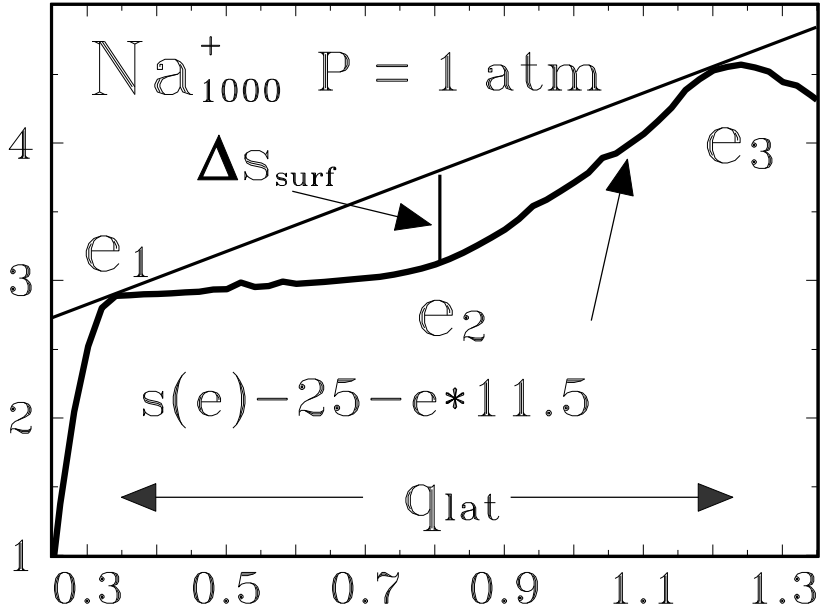}\\~\\
\caption{Micro-canonical entropy as function of excitation energy of 
  $1000$ sodium atoms at an external pressure $P=(\partial
  S(E,V)/\partial V)/(\partial S/ \partial E)=1$ atm. At $e_1$ the
  system is in the liquid phase and at $e_3$ in the pure gas phase.
  $\Delta s_{surf}$ gives the entropy-loss per atom due to surface
  correlations}\label{surftens}
\end{figure}

It is linked to the inter-phase surface tension. This is shown in a
simulation of $1000$ sodium atoms at constant external pressure
$P=1$atm.  In figure (\ref{surftens}) I show the micro-canonical
entropy $s(e)$ for $1000$ sodium atoms with realistic interactions.
Details of this calculation are given in \cite{gross157}. Here only a
few remarks: The calculations were done at a constant volume $V(E)$
which was chosen for the whole ensemble at the given energy so that
the pressure $P=\frac{\partial S}{\partial V}/\frac{\partial
  S}{\partial E}$ is given to be $1$ atm. The important and
characteristic difference to Andersen's constant pressure
ensemble\cite{andersen80} should be noticed! This is discussed in my
book that will be published soon \cite{gross174}.

The table (\ref{sodium}) gives the 4 characteristic parameters
classifying the transition of $N=200\cdots 3000$ Na-atoms at external
pressure of $1$ atm. $T_{tr}$ is the transition temperature
($T=(\partial S/\partial E)^{-1}$) in Kelvin, $q_{lat}=e_3-e_1$ is the
the latent energy per atom, $s_{boil}$ is the entropy gain of a single
atom when converted from the liquid phase into the gas phase. $\Delta
s_{surf}$ is the entropy-loss per atom due to surface correlations,
$N_{eff}$ is the average number of surface atoms of all coexisting
clusters, and $\sigma$ is the surface tension per surface atom. These
values are compared to their corresponding values of bulk sodium.
\begin{table}
\cent{
\begin{tabular} {|c|c|c|c|c|c|} 
\hline &$N_0$&$200$&$1000$&$3000$&\vecb{bulk}\\ 
\hline \hline  &$T_{tr} \;[K]$&$940$&$990$&$1095$&\vecb{1156}\\ 
\cline{2-6} &$q_{lat} \;[eV]$&$0.82$&$0.91$&$0.94$&\vecb{0.923}\\ 
\cline{2-6} {\bf Na}&$s_{boil}$&$10.1$&$10.7$&$9.9$&\vecb{9.267}\\ 
\cline{2-6} &$\Delta s_{surf}$&$0.55$&$0.56$&$0.45$&\\ 
\cline{2-6} &$N_{eff}^{2/3}$&$39.94$&$98.53$&$186.6$&$\vecbm{\infty}$\\
\cline{2-6} &$\sigma/T_{tr}$&$2.75$&$5.68$&$7.07$&\vecb{7.41}\\ 
\end{tabular} }
\caption{Transition parameters for $200\to 3000$ sodium atoms at external
pressure of $1$ atm.}\label{sodium}
\end{table}
\section{The information lost in the grand-canonical ensemble
\label{lostinfo}}
In this chapter I explain how and which part of the micro-canonical
phase diagram is lost in the conventional canonical treatment.
\begin{figure}
\includegraphics*[bb = 0 0 290 280, angle=-0, width=13cm,
  clip=true]{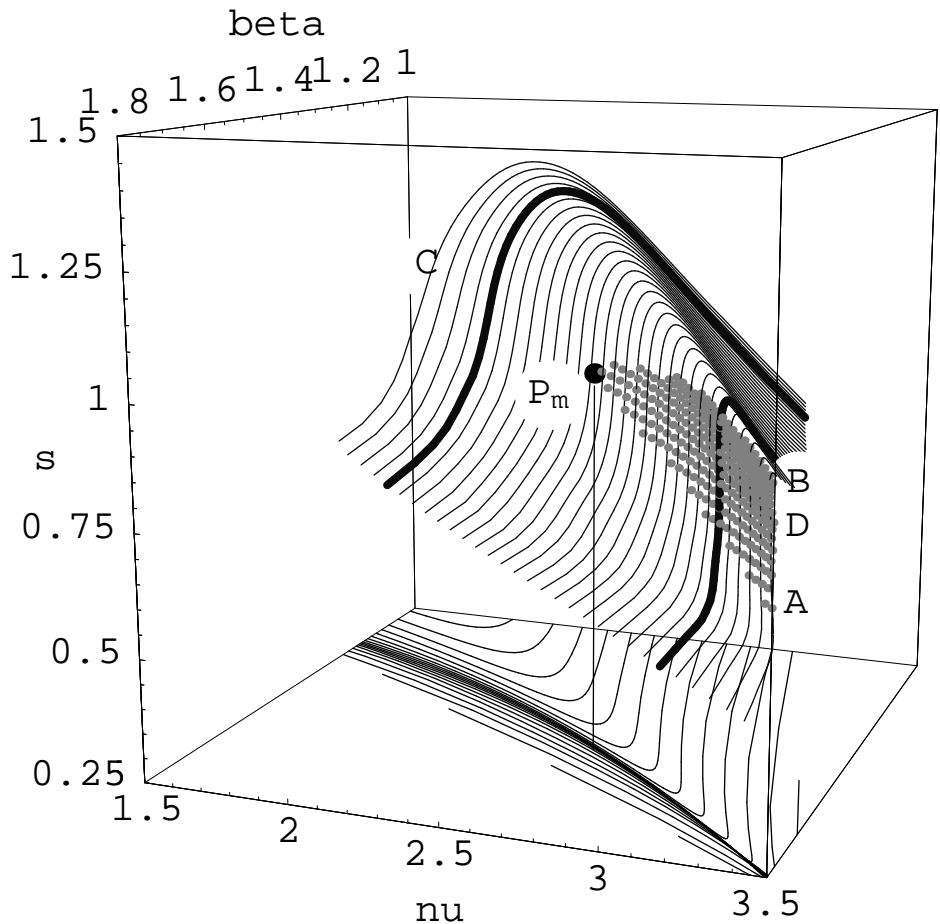}
\caption{Plot of the micro-canonical entropy 
  $s_{micro}(\beta,\nu)$ as function of the ``intensive'' variables
  ($\nu=-\beta\mu$) in the figure labeled as ``nu'' and $\beta$ is
  called ``beta''.  The lines which build the surface are lines for
  $\beta=$const. The two bold ones indicate the cuts shown in
  figs.\protect\ref{sbetaless} and \protect\ref{sbetahigh}.  The
  positions of the points $A$,$D$,$B$,$C$ defined in figure
  (\ref{det}) are only roughly indicated. The convex intruder, the
  region in figure(\ref{det}) below $\widehat{AP_mB}$ where we have
  the separation of phases and where $s_{micro}(\beta,\nu)$ becomes
  multi-valued as function of $\nu>\nu_{P_m}$ and $\beta>\beta_{P_m}$
  is indicated by shadowing. At the bottom the projection of the
  entropy surface onto the \{$\beta, \nu$\} plane is shown as contour
  plot. This would be all that could be seen in the conventional
  canonical phase-diagram. The convex part (region of
  phase-separation) is hidden behind the dark ``critical'' line.
\label{Sintens}}
\end{figure}
Figure (\ref{Sintens}) explains what happens if one plots the entropy
$s$ vs. the ``intensive'' quantities $\beta=\partial S/\partial E$ and
$\nu =\partial S/\partial N$ as one would do for the grand-canonical
ensemble: As there are several points $E_i,N_i$ with identical
$\beta,\nu$, $s_{micro}(\beta,\nu)$ is a multivalued function of
$\beta,\nu$. Here the entropy surface $s_{micro}(e,n)$ is folded onto
itself. In the projection in fig.\ref{Sintens}, these points show up
as a black critical line (dense region). Here this black line continues
over the multi-critical point $P_m$ towards $C$ indicating the
direction to the critical point of the ordinary $q=3$ Potts model at
$n=1$ (zero vacancies). Between $P_m$ and $C$ the slopes
\begin{eqnarray}
\left.\frac{\partial s}{\partial \beta}\right|_\nu&=&
\frac{1}{d}[\beta s_{nn}-\nu s_{ne}]\\
\mbox{or}&&\nonumber\\
\left.\frac{\partial s}{\partial \nu}\right|_\beta&=&
-\frac{1}{d}[\beta s_{en}-\nu s_{ee}]
\end{eqnarray}
are negative large but finite.

The information given by the projection would be all information which
can be obtained from the conventional grand-canonical entropy
$s(T,\nu,V)$, if we would have calculated it from the Laplace
transform, eq.(\ref{grandsum}). The shaded region will be lost.

The upper part of figure (\ref{Sintens}) shows $s_{micro}(\beta,\nu)$
in a three dimensional plot. The lines building the entropy surface
are lines of equal $\beta$. The images of the points $A,D,B,C$ defined
in fig.\ref{det} are roughly indicated.  The back folded branches, the
convex intruder of $s(e,n)$ between the lines $\widehat{AP_mB}$ and
$\widehat{ADB}$, the region of phase separation, can here be seen from
the side (shadowed). It is jumped over in eq. (\ref{grandsum}) and
gets consequently lost in $Z(T,\nu)$. This demonstrates the far more
detailed insight into phase transitions and critical phenomena
obtainable by micro-canonical thermo-statistics not accessible to the
canonical treatment, c.f. the similar arguments of Gibbs
\cite{gibbs06}.

\begin{figure}
  \includegraphics*[bb = 0 0 290 180, angle=-0, width=10cm,
  clip=true]{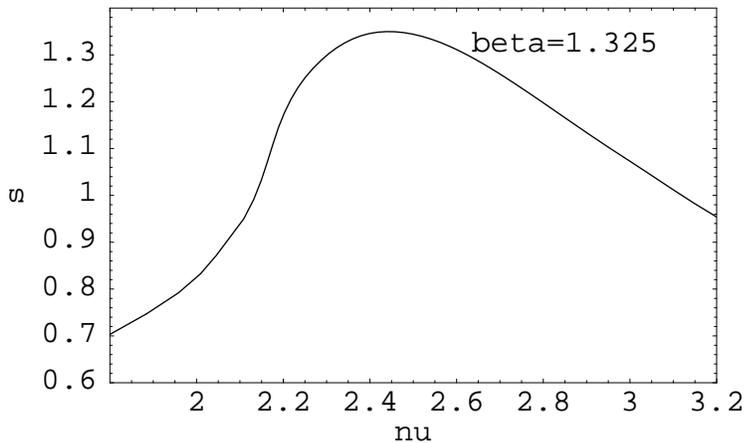}
\caption{Plot of the entropy $s(\beta=1.325,\nu)$\label{sbetaless}}
\end{figure}

\begin{figure}
  \includegraphics*[bb = 0 0 290 180, angle=-0, width=10cm,
  clip=true]{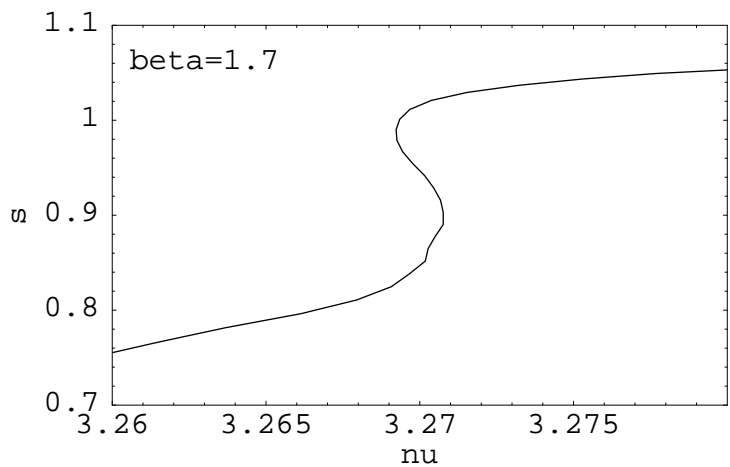}
\caption{Plot of the entropy $s(\beta=1.7,\nu)$\label{sbetahigh}}
\end{figure}

In the next two figures the cross-section through $s(\beta,\nu)$ at
constant $\beta$ along the bold lines from fig.(\ref{Sintens}) is shown
in figure (\ref{sbetaless}) below the multi-critical point
$\beta_m=1.48$ and in figure (\ref{sbetahigh}) above it.  The latter
clearly shows the back-bending of $s(\beta,\nu)$.
\section{Convex entropy --- Violation of the Second Law ?}\label{secondlaw}
At this point it is worth-wile to spend some words on a popular
misunderstanding connected with the eventual convexity of the entropy
as function of ``extensive'' quantities like the energy: The convex
parts of $S(E,N)$ violate van Hove's concavity condition
\cite{vanhove49,hill55}.

One may believe that this is also a contradiction to the second law
of thermodynamics: At a convex region of $S(E,N)$ a split of the
system into two pieces with entropies $S_1(E_1,N_1)$ and
$S_2(E_2,N_2)$ would have
$S_1(E_1,N_1)+S_2(E_2,N_2)>S(E_1+E_2,N_1+N_2)$. So the system seems to
{\em gain} entropy by splitting.

This, however, is an error. The Boltzmann entropy as defined in
eq.(\ref{boltzm}) is already the logarithm of the sum over {\em all}
possible configurations of the system at the given energy. The split
ones are a subset of these. Their partial phase space $W_{split}$ is
of course $\le$ the total $W$. The entropy $S_{split}=ln(W_{split})$
is $\le$ the total entropy. Evidently, the split system looses some
surface entropy $S_{surf}$ at the separation boundary due to
additional correlations imposed on the particles at the boundary.  The
entropy after split is consequently:
\begin{eqnarray}
S_{split}&=&S_1(E_1,N_1)+S_2(E_2,N_2)-S_{surf}\nonumber\\
&\le&S(E_1+E_2,N_1+N_2),
\end{eqnarray}
It is a typical finite size effect. $S_{surf}/V$ vanishes in the limit
$V\to\infty$ for interactions with {\em finite} range. The entropy is
non-extensive for finite systems but becomes extensive in the limit,
and van Hove's theorem \cite{vanhove49} is fulfilled. This is of
course only under the condition that $\lim_{V\to\infty}S_{surf}/V =0$.
So in the case of a self-gravitating system the convex intruder and
the negative specific heat will not disappear
\cite{hertel71,hertel72,stahl95,kiessling95,laliena98}.

In general this is of course a trivial conclusion: An additional
constraint like an artificial cut of the system can only reduce phase
space and entropy. The Second Law is automatically satisfied in the
Boltzmann formalism whether $S$ is concave or not, whether $S$ is
``extensive'' or not.

A positive (wrong) curvature introduces problems to the geometrical
interpretation of thermodynamics as formulated by Weinhold
\cite{weinhold75,weinhold78} which relies on the non-convexity of
$S(E,N)$.  Weinhold introduces a metric like
\begin{eqnarray}
g_{ik}&=&-\frac{\partial^2S}{\partial X^i\partial X^k}\\
&&\mbox{where we identify :}\nonumber\\
X^1&=&E\nonumber\\
X^2&=&N .\nonumber\\
&&\mbox{The thermodynamic distance is defined as :}\nonumber\\
\Delta_{a,b}&=& \sqrt{[X^i(a)-X^i(b)]g_{ik}[X^k(a)-X^k(b)]}.
\end{eqnarray} 
Evidently, a negative metric $g_{ik}$ is here not allowed. Of course
Weinhold's theory does not apply to finite systems with phase
transitions.
\section{Conclusion} 
Micro-canonical thermo-statistics describes how the entropy $s(e,n)$
as defined entirely in mechanical terms by Boltzmann depends on the
conserved ``extensive'' variables: energy $e$, particle number $n$,
angular momentum $L$ etc. It is well defined for finite systems
without invoking the thermodynamic limit.  Thus in contrast to the
conventional theory, we can study phase transitions also in ``small''
systems or other non-extensive systems. In this simulation we could
classify phase transitions in a ``small'' system by the topological
properties of the determinant of curvatures $d(e,n)$,
eq.(\ref{curvdet}) of the micro-canonical entropy-surface $s(e,n)$:

In the micro-canonical ensemble of a ``small'', non-extensive system,
phase transitions are classified unambiguously by the following
topology of the curvature determinant:
\begin{itemize}
\item A {\bf single stable} phase by $d(e,n)>0$ ($\lambda_1<0$).
Here $S(E,N)$ is concave in both directions.
Then there is a one to one mapping of canonical\lra micro-ensemble.
\cent{\includegraphics*[bb = 117 13 493 646, angle=-90, width=9cm,  
clip=true]{concave.eps}}
Then the last two terms in
\begin{equation}
\frac{F(T,\mu,V)}{V}\to e_s-\mu n_s-Ts_s+T
\frac{\ln{(\sqrt{d(e_s,n_s)})}}{V}+o(\frac{\ln{V}}{V})\label{asympt}
\end{equation}
can be neglected for large volume.
\item A {\bf transition of first order} with phase separation and
  surface tension is indicated by $d(e,n)<0$ ($\lambda_1>0$). $S(E,N)$
  has a convex intruder in the direction
  {\boldmath$\vecbm{v}_{\lambda_1}$} of the largest curvature. Because of 
$d\le0$ the second last term in eq.(\ref{asympt}) is complex or diverges.
  \cent{\includegraphics*[bb = 130 8 494 646, angle=-90, width=9cm,
    clip=true]{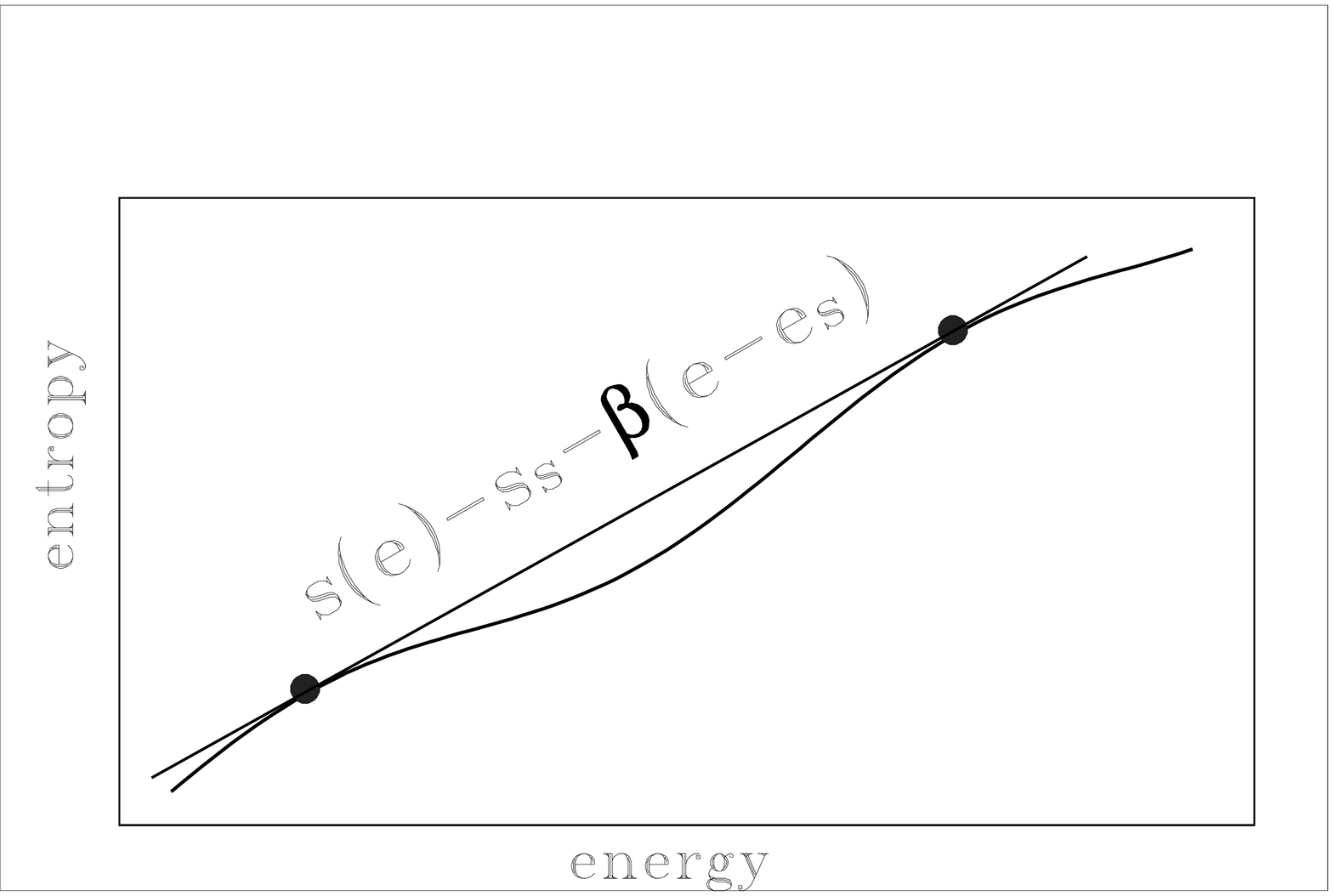}} The whole convex area of \{e,n\} is mapped
  into a single point in the
  canonical ensemble.\label{convex}\\
  I.e. if the curvature of $s(e,n)$ is $\lambda_1\ge 0$ {\bf both
    ensembles are not equivalent.}
\item A {\bf continuous (``second order'')} transition with vanishing  
surface tension, where two neighboring phases become  
indistinguishable, is indicated by lines (critical) with
$d(e,n)=0$ and {\boldmath$\vecbm{v}_{\lambda=0}\cdot\vecbm{\nabla}d=0$}.
These are the catastrophes of the Laplace transform $E\to T$
\item Finally a {\bf multi-critical point} where more than two phases 
become  indistinguishable is at the branching of several lines with
$d=0$, {\boldmath$\vecbm{\nabla} d=0$}.
\end{itemize}

Our classification of phase transitions by the topological structure
of the micro-canonical Boltzmann entropy $s(e,n)$ is close to the
natural experimental way to identify phase transitions of first order
by the inhomogeneities at phase separation boundaries. This is
possible because the micro-canonical ensemble does not suppress
inhomogeneities in contrast to the grand-canonical one, as was
emphasized already by Gibbs \cite{gibbs06}.  Inter-phase boundaries are
reflected in ``small'' systems by a convex intruder in the entropy
surface. With this extension of the definition of phase transitions
to ``small'' systems there are remarkable similarities with the
transitions of the bulk.  Moreover, this definition agrees with the
conventional definition in the thermodynamic limit (of course, in the
thermodynamic limit the largest curvature $\lambda_1$ approaches $0$
from above at phase transitions of first order). The region of phase
separation remains inaccessible in the conventional grand-canonical
ensemble.

We believe, however, that the various kind of transitions discussed
here have their immediate meaning in ``small'' and non-extensive
systems independently whether they are the same in the thermodynamic
limit (if this then exist) or not. For systems like the Potts model
that have a thermodynamic limit it might well be possible that the
character of the transition changes towards larger system size.

The great conceptual clarity of micro-canonical thermo-statistics
compared to the grand-canonical one is clearly demonstrated.  Not only
that, we showed that the micro-canonical statistics gives more
information about the thermodynamic behavior and more insight into
the mechanism of phase transitions than the canonical ensemble: About
half of the whole \{E,N\} space, the intruder of $S(E,N)$ or the
region between the ground state and the line $\widehat{AP_mB}$ in
figure (\ref{det}), gets lost in conventional grand-canonical
thermodynamics.  Without any doubts this contains the most
sophisticated and interesting physics of this system. We emphasized
this point already in \cite{gross170} there, however, with still
limited precision. Due to our refined simulation method
\cite{gross173} this could be demonstrated here with uniformly good
precision in the whole \{$E,N$\} plane. It turns out that not only are
non-extensive systems a new and rich realm for thermodynamics but
moreover {\em non-extensivity makes phase transitions much more
  transparant} which is no surprise as phase transitions of first
order are coupled to situations where a system {\em prefers to become
  inhomogeneous, i.e. non-extensive}.

Finally, we should mention that micro-canonical thermo-statistics
allowed us to compute phase transitions and especially the surface
tension in realistic systems like small metal clusters
\cite{gross157}. Our finding \cite{gross173} clearly disproves the
pessimistic judgement by Schr\"odinger \cite{schroedinger44} who
thought that Boltzmann's entropy is only usefull for gases.  A recent
application of micro-canonical thermo-statistics to thermodynamically
unstable, collapsing systems under high angular momentum is found at
\cite{laliena98}.
\\~\\
\noindent Acknowledgment: First if all I thank to E.V.Votyakov
for performing most of the numerical work.  I thank M.E.Fisher for the
suggestion to study the Potts-3 model and to test how the
multicritical point is described micro-canonically.

\end{document}